\documentclass[12pt]{article}
\usepackage[authoryear]{natbib}
\usepackage{array,epsfig,fancyheadings,rotating}
\usepackage{amsmath}
\usepackage{graphicx}
 \usepackage{afterpage}
\usepackage{pdflscape}
\usepackage[]{lineno}
\usepackage{amsmath}
\usepackage{amssymb}
\usepackage{amsfonts}
\usepackage{multirow}
\usepackage{amsthm}
\usepackage{graphicx}
\usepackage{subfig}
\usepackage{url} 

\newcommand{\blind}{0}

\graphicspath{{./figures/}}
\newcommand{\bs}[1]{\boldsymbol{#1}}
\begin{document}

\def\spacingset#1{\renewcommand{\baselinestretch}%
{#1}\small\normalsize} \spacingset{1}


\if0\blind
{
  \title{\bf Fast expectation-maximization algorithms for spatial generalized linear mixed models}
  \author{Yawen Guan$ ^1 $\thanks{Email: yawen.guan@unl.edu} and Murali Haran$ ^2 $ \thanks{Email: mharan@stat.psu.edu} \hspace{.2cm}\\
    $ ^1 $\small{Department of Statistics, University of Nebraska, Lincoln, Nebraska, United States}\\
    $ ^2 $\small{Department of Statistics, Pennsylvania State University, University Park, United States}
    }
  \maketitle
} \fi

\if1\blind
{
  \bigskip
  \bigskip
  \bigskip
  \begin{center}
    {\LARGE\bf Fast expectation-maximization algorithms for spatial generalized linear mixed models}
\end{center}
  \medskip
} \fi

\bigskip
\begin{abstract}
Spatial generalized linear mixed models (SGLMMs) are popular and flexible models for non-Gaussian spatial data. They are useful for spatial interpolations as well as for fitting regression models that account for spatial dependence, and are commonly used in many disciplines such as epidemiology,  atmospheric science, and sociology. Inference for SGLMMs is typically carried out under the Bayesian framework at least in part because computational issues make maximum likelihood estimation challenging, especially when high-dimensional spatial data are involved. Here we provide a computationally efficient projection-based maximum likelihood approach and two computationally efficient algorithms for routinely fitting SGLMMs. The two algorithms proposed are both variants of expectation maximization algorithm, using either Markov chain Monte Carlo or a Laplace approximation for the conditional expectation. Our methodology is general and applies to both discrete-domain (Gaussian Markov random field) as well as continuous-domain (Gaussian process) spatial models. We show, via simulation and real data applications, that our methods perform well both in terms of parameter estimation as well as prediction. Crucially, our methodology is computationally efficient and scales well with the size of the data and is applicable to problems where maximum likelihood estimation was previously infeasible.
\end{abstract}

\noindent%

\vfill

\newpage
\spacingset{1.7} 
\section{Introduction}

Non-Gaussian spatial data arise in a number of disciplines, for instance when modeling disease incidence in epidemiology \citep[see, for example][]{Diggle1998,HughesHaran} or modeling weed counts and plant disease in agriculture \citep{Christensen2002,Haozhang2002}. Spatial generalized linear mixed models (SGLMMs) are convenient and flexible models for such data. Following two seminal papers, \cite{Diggle1998} and \cite{Besag1991}, SGLMMs have been very popular, not only in mainstream statistics but also in many other disciplines. These models are useful both for data observed on a continuous spatial domain such as at irregularly-positioned sampling locations and data observed on a discrete spatial domain such as county-level data. In this manuscript, we propose two fast maximum likelihood (ML) inference algorithms for a projection-based approach that are applicable for both the continuous and discrete spatial domains. 

Inference for SGLMMs is commonly carried out under the Bayesian paradigm \citep[see][]{banerjee2014hierarchical,Muralibook}. However, constructing efficient Markov chain Monte Carlo (MCMC) samplers for fitting such models to large data sets is often challenging. There are two major computational challenges: (1) computational issues due to high-dimensional random effects that are typically heavily correlated among themselves (cross-correlated) -- these often result in slow mixing MCMC algorithms; (2) expensive calculations involving large matrices. An additional issue is spatial confounding between fixed and random effects -- this can result in slow mixing and problems with parameter interpretation \citep[cf.][]{GuanHaran2016,RHZ2006,ephraim2015,HughesHaran}. Under a Bayesian framework, the high-dimensional computational challenges for SGLMMs have been addressed via the predictive process approach \citep{Banerjee2008} and the Vecchia-Laplace approximation \citep{Zilber2019}, the MCMC mixing issues have been addressed by various reparameterizations \citep[cf.][]{RobustMCMC2006, Haran2003,rue2005gaussian}, and the confounding issues have been addressed in \cite{RHZ2006}. \cite{RueINLA} provided a fast inferential approach based on nested Laplace approximations and \cite{Link2011} suggested how this approximation may be adapted to continuous spatial domain SGLMMs. Recently, via projection-based methods, \cite{HughesHaran} and \cite{GuanHaran2016} have addressed both the above computational as well as confounding issues within a Bayesian approach. 

We consider ML inference for SGLMMs, which had received less attention, at least in part, because of computational challenges. For data sets with just a few hundred data points, a Monte Carlo expectation-maximization (MCEM) algorithm \citep{Haozhang2002} and a Monte Carlo maximum likelihood (MCML) algorithm \citep{Christensen2004} were proposed for fitting SGLMMs. However, neither algorithm extends easily to large data sets because they both require simulation of the high-dimensional latent variables, which is computationally expensive when the data sets are large. \cite{Sengupta2013b,sengupta2013} developed fast ML inference for large non-Gaussian observations by approximating the spatial random effects with basis functions to resolve computational issues. The projection-based methods used in this manuscript \citep{HughesHaran,GuanHaran2016} can be thought of as a fixed-rank approach, but use data driven basis functions. More recently, \cite{ParkHaran2020} develop a MCML algorithm for fitting SGLMMs. This approach combines \cite{Christensen2004} and projection approach to handle higher dimensional problems than previously considered.
However, MCML algorithms can be challenging to implement for non-experts. The expectation-maximization (EM) algorithms we 
propose in this paper are generally easier to implement than MCML. 
Furthermore, after the seminal paper by Zhang (2002) on EM for SGLMMs, there has been relatively little if any work on EM algorithms for the kind of 
generalized linear models with high-dimensional dependent latent variables that we consider in this manuscript. To our knowledge, the algorithms 
here are therefore among the first viable EM algorithms for such models for large data sets: our approach for using a projection-based dimension 
reduction of the latent variables  opens up interesting new avenues for developing EM algorithms that are practical for such models. \cite{Bonat2016} developed an approximate likelihood-based approach, which substitutes a Laplace approximation (LA) for MC simulation. However, it is unclear how well this approach will work for high-dimensional problems as it requires Gaussian approximations to the full conditional distribution of the high-dimensional latent variable. 

Our contribution in this manuscript is to provide computationally efficient ML inference for SGLMMs for large data with the ability to address the computational issues arise from spatial confounding. We develop two variants of the EM algorithm, Markov chain Monte Carlo EM (MCEM) and Laplace approximation EM (LAEM), for maximum likelihood estimation. Our approach provides the ability to fit SGLMMs routinely by (i) having an automated algorithm for estimation, (ii) reducing the computational cost of the estimation algorithm, (iii) addressing computational issues arise from spatial confounding, and (iv) sidestepping the need to provide hyper-priors for parameters about which there is often little available information. Our manuscript also contributes to the study of practical issues in constructing MCEM algorithms in the context of a challenging latent variable model. We believe, as applied statisticians ourselves, that the above characteristics are useful to researchers who use SGLMMs in applications. For problems that involve fitting an SGLMM to a spatial data set in more complicated settings where an additional hierarchy in the modeling framework becomes necessary, for instance where multiple data sets need to be integrated, we would likely revert to a Bayesian approach. 

The outline of the remainder of the paper is as follows. In Section \ref{Sec:sec2}, we describe SGLMMs and spatial confounding. We introduce in Section \ref{Sec:proj} the projection-based SGLMMs and in Section \ref{Sec:MLInference} the MCEM and LAEM algorithms for ML inference. We study our method via a simulation study in Section \ref{Sec:simulation} and apply it to two data sets in Section \ref{Sec:application}. We conclude with a discussion and potential areas for future work in Section \ref{chap3:sec7}.

\section{Spatial Generalized Linear Mixed Models}\label{Sec:sec2}
\subsection{Models}
SGLMMs provide a framework for analyzing spatially dependent non-Gaussian observations. Let $ Z(\bs{s}) $ denote the response variable, $ \bs{x}(\bs{s}) = (x_1(\bs{s}),...,x_p(\bs{s}))^T$ denote the explanatory variables, and $ W(\bs{s}) $ represent a spatial random field, where $ \bs{s} \in \mathbb{R}^2 $ indicates a spatial location. For data obtained at a finite collection of locations $\mathcal{S} = \left\lbrace \bs{s}_1,\dots,\bs{s}_n\right\rbrace $, we write $ Z_i = Z(\bs{s}_i) $ and let $ \bs{Z} = (Z_1,...,Z_n)^T $ be a vector of the observed response variable at $ \mathcal{S} $. Similarly, let $ X = (\bs{x}_1,...,\bs{x}_n) $ and $ \bs{W} = (W_1,\dots,W_n)^T $ be the corresponding finite counterparts. The SGLMMs can be defined with three components.

\begin{itemize}

\item[(i)] A model that captures spatial dependence. This can change depending  on whether the data are on a discrete (lattice) or continuous spatial domain. 

For a continuous spatial domain, $ W(\bs{s})$ is often modeled as a zero-mean stationary Gaussian random field with $ \text{cov}(W(\bs{s}),W(\bs{s}'))$ $= C(||\bs{s}-\bs{s}'||) $ for $ \bs{s},\bs{s}' \in \mathcal{R}^2 $, where the covariance function $ C(\cdot) $ depends on a vector of parameters $ \bs{\theta} $. Hence, $ \bs{W} $ follows a multivariate normal distribution, $$	f(\bs{W}|\bs{\theta}) \propto \vert\Sigma_{\bs{\theta}}\vert^{-1/2} \exp \left( -\frac{1}{2} \bs{W}^T\Sigma_{\bs{\theta}}^{-1}\bs{W}  \right).$$ A frequently used covariance function, assuming stationarity and isotropy, is the Mat\'{e}rn class \citep{Stein1999}. 

For a discrete spatial domain, $ \bs{W} $ is typically modeled as a zero-mean Markov random field. The index of $ W_i $ indicates a node on a lattice, typically denoting a geographic block. The neighboring structure among blocks is defined through an $n \times n$ adjacency matrix $ A $, with $\text{diag}(A) = 0$ and $A_{ij} = 1$ if the $i^{th}$ and $j^{th}$ locations are connected \citep{Besag1991}. A popular model for $\bs{W}$ is the intrinsic conditionally auto-regressive (ICAR) model, $$ f(\bs{W}|\tau) \propto \tau^{\text{rank} (Q)/2} \exp \left( -\frac{\tau}{2} \bs{W}^TQ\bs{W}\right),$$ where $\tau$ is a  parameter that controls the smoothness of the spatial field and $Q = \text{diag}(A\bs{1}) -A$ is the precision matrix, and $ \bs{1} $ is an $ n $-dimensional vector of ones.

\item[(ii)] Conditional on $ \bs{W} $ and $ \bs{\beta} $, observations $ \bs{Z} $ are independently distributed with distribution function $ \prod_{i=1}^{n}f_{Z_i\vert W_i}(Z_i| W_i,\bs{\beta}) $. Each observation has a site-specific conditional mean $  \mu_i = E\left[Z_i | W_i,\bs{\beta}\right]$.

\item[(iii)] A link function $ g $ that relates the conditional mean to a linear model, $g\left( \mu_i\right)  = \bs{x}_i^T\bs{\beta} + W_i$. For instance, it is common to use a log link for counts. 
 
\end{itemize}

In the remaining sections, we use $  \bs{\theta} $ to denote parameters of the spatial random fields for both continuous and discrete cases to unify the notations, although $  \bs{\theta} $ is a scalar in the latter case. The observed-data likelihood or SGLMMs has the form
\begin{equation}\label{eqn:fullL}
L(\bs{\beta},\bs{\theta} ;\bs{Z}) = \int_{\mathbb{R}^n} \left\lbrace \prod_{i = 1}^{n} f_{Z_i|W_i} \left( Z_i| W_i,\bs{\beta} \right)\right\rbrace  f_{W} \left( \bs{W}|\bs{\theta}\right) d\bs{W},
\end{equation}
which involves a high-dimensional integral and is typically not available in closed form. Therefore, direct maximization of (\ref{eqn:fullL}) is infeasible. MCML (Geyer and Thompson 1992) and MC versions of EM algorithms (cf., Wei and Tanner 1990, McCulloch 1994) have been proposed to approximate the integration in (1) using MC samples \citep{Christensen2004, Haozhang2002}. These MC methods require simulations from the conditional distribution of random effects given the data, $ f_{W} \left( \bs{W}|\bs{\theta}, \bs{Z} \right) $, for both inference and prediction. These methods work quite well for data sets that are relatively small, say in the hundreds. When confronted with thousands of data points or more, these methods become computationally challenging. This is largely because, like in the Bayesian approach, the number of random effects grows with the size of the data. This results in a high-dimensional integration problem at each iteration of the EM algorithm which, in turn, leads to an unstable MCEM algorithm. Furthermore, it becomes difficult to construct a fast mixing MCMC algorithm at each expectation step because the random effects are often highly cross-correlated. In addition to addressing these challenges via the projection-based approach, we provide some guidance on how to tune the algorithm, including, for instance, how to determine appropriate MC sample sizes for each step of the algorithm. 

\subsection{Spatial Confounding}
Let $ P_{[X]} = X(X^TX)^{-1}X^T$ and $ P_{[X]}^{\perp} = I - P_{[X]}$ denote the orthogonal projections onto the span of X and its complement, respectively. The confounding problem therefore arises in much the same way as in multicolinearity problems with standard regression models. The only difference here is that the confounding arises due to the spatial random effects. The linear model for site-specific conditional means, $ \bs{\mu} = (\mu_1,...\mu_n)^T $, is $  g\left(\bs{\mu}\right) = X\bs{\beta}+ \bs{W} = X\bs{\beta}+P_{[X]}\bs{W} +P_{[X]}^{\perp}\bs{W}$. Since $P_{[X]}\bs{W} $ is confounded with $ X $, \cite{Hodges2010} suggested that it should be removed from the model to alleviate spatial confounding. However, \cite{ephraim2015} argues that when $P_{[X]}\bs{W}$ is ``removed" from the model, its effect is combined with $\bs{\beta}$ and an \textit{a posteriori} adjustment should be performed to obtain valid inference about $\bs{\beta}$. This way of restricting the random effects to be orthogonal to the fixed effects is also called restricted spatial regression (RSR) model. Methods for addressing these problems have been developed and studied for both continuous and discrete domain data \citep[cf.][]{RHZ2006, ephraim2015, GuanHaran2016, HughesHaran}. 

Recent work also address spatial confounding issue from a causal prospective, where the main goal is to draw inference for the effect of spatially observed exposures under missing confounders that posit spatial structures. A number of assumptions are required for making valid causal interpretations, readers who are interested may refer to a recent review paper \cite{reviewspcausal} and references therein.

In our experience, RSR models often provide parameter estimates that are very close to the true values, but the interval estimates are typically very narrow, therefore, leading to a higher type-I error as noted by \cite{ephraim2015}. However, RSR model eliminates the colinearity between fixed and random effects, which greatly improves the mixing problem in MCMC sampling and the latent random effect estimation. In this manuscript, the random projection approach (introduced in Section 3) does not address the confounding problems, but rather the computational issues due to confounding and provides an alternative method to fit the restricted models proposed by \cite{RHZ2006}. We also propose to use parametric bootstrap (Section 4.1) for obtaining interval estimates which were shown to have better coverages.

\section{A Projection-Based Approach to Dimension Reduction}\label{Sec:proj} 
We consider two projection-based models for the continuous and discrete spatial domains \citep{GuanHaran2016, HughesHaran}. Both models leverage efficient reparameterizations to (1) reduce the dimension of the random effects and (2) alleviate spatial confounding. They share a common form, $P^\perp_{[X]}\bs{W} \approx M\bs{\delta}$, where $ \bs{\delta}$ is an $ m -$dimensional vector with nearly independent elements and $ M $ is an $ n\times m $ projection matrix that preserves the spatial information of $\bs{W}$. The projection matrix for the continuous case is computed based on the covariance matrix driven by the data, while for the discrete case it is based on the graph based on the neighboring structure. 

For the continuous case, an example is $ C(h) = \sigma^2(1+\sqrt{3}h/\phi)\exp(-\sqrt{3}h/\phi) $, which corresponds to the Mat\'{e}rn covariance model with smoothness $\nu = 1.5$ and $  \bs{\theta} = (\sigma^2,\phi)^T $. Let $ R_\phi $ denote the correlation matrix of $ \bs{W} $ and $ \Sigma_{ \bs{\theta}} = \sigma^2 R_\phi $. \cite{GuanHaran2016} proposed to reparameterize $ \bs{W} $ using the first $ m (<<n) $ principal component of $ R_\phi$ and then project the reduced-dimensional random effects to the orthogonal span of $ X $. Let $ U_\phi = \left[\bs{u}_1,...,\bs{u}_m\right]$ denote the first $ m $ eigenvectors and $ D_\phi =\text{diag}(\lambda_1,\dots,\lambda_m)$ a diagonal matrix containing eigenvalues of $ R_\phi $. The reparameterized random effects have the form $\widetilde{\bs{W}} = U_\phi D_\phi^{1/2}\bs{\delta}$, which result in independent random effects $ \bs{\delta}|\sigma^2,\phi \sim N(0,\sigma^2 I)$, and $P^{\perp}_{[X]}\widetilde{\bs{W}} = M_\phi\bs{\delta} $, where $M_\phi = P^{\perp}_{[X]}U_\phi D_\phi^{1/2}$, is restricted to be orthogonal to the fixed effects. The hierarchical model becomes 
\begin{equation*} 
g\left\lbrace E\left[\bs{Z}|\bs{\beta},M_\phi,\bs{\delta}\right]\right\rbrace = X\bs{\beta} + M_\phi\bs{\delta},  \hspace{1em}
\bs{\delta}|\sigma^2,\phi {\sim} \text{N}(\bs{0},\sigma^2I).
\end{equation*}

If exact eigendecomposition is computationally infeasible, say when there are several thousands of data points, we can   approximate it using a probabilistic version of the Nystr\"om's method \citep{drineas2005nystrom}. An outline of the approximation algorithm is presented in the supplementary materials; details are provided in \cite{GuanHaran2016,A.Banerjee2012}. 

For the discrete case, the reparameterization is based on the first $ m $ principal components, $M_A$, of the Moran operator $P^{\perp} AP^{\perp}$ \citep{HughesHaran}. The model has the form $g\left\lbrace E\left[\bs{Z}|\bs{\beta},\bs{\delta} \right]\right\rbrace  = X\bs{\beta} + M_A\bs{\delta}, \hspace{0.5em}
p(\bs{\delta}|\tau) \propto \tau^{m/2} \exp \left( -\frac{\tau}{2} \bs{\delta}^TQ_\delta\bs{\delta}  \right), \text{where } Q_\delta = M_A^TQM_A.$

\section{ML Inference Methods}\label{Sec:MLInference}
Two variants of the EM algorithm are derived here for fitting the projection-based models. The EM algorithm iterates between the expectation step (E-step) and maximization step (M-step) for parameter estimation. The two EM variants proposed here are distinct in their approximations to the conditional expectation in E-step; one uses MC averages and the other uses LA. 

The projection-based model facilitates fast ML inference because  its observed-data likelihood has a much smaller dimension integration compared to the full model \eqref{eqn:fullL},  
\begin{equation}\label{eqn:projL}
L(\bs{\beta},\bs{\theta};\bs{Z}) = \int_{\mathbb{R}^m} \left\lbrace \prod_{i = 1}^{n} f_{Z_i|M\bs{\delta}} \left( Z_i|M\bs{\delta},\bs{\beta} \right)\right\rbrace  f_{\bs{\delta}} \left( \bs{\delta}|\bs{\theta}\right) d\bs{\delta}.
\end{equation} 
For instance, in our simulation study $ m = 50 $ is sufficient for a data size of 1,000 in some settings, based on the rank selection guidelines provided in Section \ref{sec:rkselection}; moreover, $ \bs{\delta} $ is less correlated than the original random effects. The reduced-dimensional and de-correlated random effects make it easier to construct a sampling algorithm (Section \ref{Sec:MCMCEM}). The reparameterization also reduces matrix operation cost for the LA (Section \ref{sec:LAML}). 

\subsection{Projection-Based EM}\label{Sec:EM}
The projection-based EM algorithm is outlined here, and details for the two proposed EM variants are presented in the subsequent sections. For ease of representation, we write $ \bs{\psi} = (\bs{\beta},\bs{\theta}) $ and let $ f_{Z,\delta}(\bs{Z},\bs{\delta}; \bs{\psi}) $ denote the integrand in (\ref{eqn:projL}). In an EM algorithm, random effects $ \bs{\delta} $ are treated as missing data and $f_{Z,\delta}(\bs{Z},\bs{\delta}; \bs{\psi})$ is called the complete-data likelihood.

Let $ \bs{\psi}^{(t)} $ be the current estimate of the ML estimator (MLE) $ \hat{\bs{\psi}} $. The EM algorithm iterates between the following two steps for $ t = 1,2,3,... $, 

\begin{itemize}
\item[] \textbf{E-step:} under the current parameter value $ \bs{\psi}^{(t)} $, compute $$	Q (\bs{\psi},\bs{\psi}^{(t)})  = E[\ln f_{Z,\delta}(\bs{Z},\bs{\delta}; \bs{\psi}) |\bs{Z},\bs{\psi}^{(t)}]$$
\item[] \textbf{M-step:} find $ \bs{\psi}^{(t+1)} $ so that $ Q(\bs{\psi}^{(t+1)},\bs{\psi}^{(t)})\ge Q(\bs{\psi}^{(t)},\bs{\psi}^{(t)})$
\end{itemize}

\noindent until the pre-specified stopping criterion is reached. One stopping rule is similar to the framework of determining MC sample sizes based on the ascent-based approach \citep{ascentMCEM}; details are in the supplementary materials. Under some regularity conditions, the EM sequence converges to the unique MLE \citep{wu1983}.

We use a gradient approach for obtaining $ \bs{\psi}^{(t+1)} $ in the M-step, where a one-step Newton-Raphson replaces the maximization. This EM gradient algorithm speeds up EM convergence and is proven to be useful in the classical settings \citep[cf.][]{lange1995gradient}; it was later extended to fitting SGLMMs \citep{Haozhang2002} for problems where the data size is relatively small. To maximize $ Q (\bs{\psi},\bs{\psi}^{(t)})$, we find its first and second derivative, $ Q^\prime $ and $ Q^{\prime\prime} $, with respect to $ \bs{\psi} $, then update the parameters using $\bs{\psi}^{(t+1)} =\bs{\psi}^{(t)}- Q^{\prime\prime}(\bs{\psi}^{(t)})^{-1} Q^{\prime}(\bs{\psi}^{(t)})$. 
%

When the derivatives, $\partial/\partial\bs{\psi}\ln  f_{Z,\delta}(\bs{Z},\bs{\delta}; \bs{\psi}) $ and   $\partial^2/\partial\bs{\psi}\partial\bs{\psi}^T\ln  f_{Z,\delta}(\bs{Z},\bs{\delta}; \bs{\psi}) $, are available in closed form, their respective conditional expectations

\begin{equation}\label{eqn:updateequations}
\begin{aligned}
Q^\prime&=E\left[\frac{\partial}{\partial\bs{\psi}}\ln f_{Z,\delta}(\bs{Z},\bs{\delta}; \bs{\psi})\vert \bs{Z},\bs{\psi}^{(t)}\right] \\
Q^{\prime\prime} &=  E\left[\frac{\partial^2}{\partial\bs{\psi}\partial\bs{\psi}^T}\ln f_{Z,\delta}(\bs{Z},\bs{\delta}; \bs{\psi})\vert \bs{Z}, \bs{\psi}^{(t)}\right],
\end{aligned}
\end{equation} 
can be approximated using MC samples or a LA. For the projection-based models in Section \ref{Sec:proj}, we have closed form expressions of the derivatives for all parameters except the range parameter (in the continuous case), and $ Q^{\prime\prime} $ is block diagonal. The latter results in separate updating equations for the regression and spatial parameters. 

Estimation for $ \bs{\beta} $ is the same for both continuous and discrete cases. If the conditional distribution of the response variable is from the exponential family, for instance, the binomial or Poisson model,  and the link function is canonical, then we have
\begin{equation} \label{eqn:dbeta} 
\begin{aligned}
\frac{\partial \ln f(\bs{Z}\vert M\bs{\delta},\bs{\beta})}{\partial \bs{\beta}} &= X^T\left(\bs{Z}-E\left[\bs{Z}\vert M\bs{\delta},\bs{\beta}\right]\right) \\
\frac{\partial^2 \ln f(\bs{Z}\vert M\bs{\delta},\bs{\beta})}{\partial\bs{\beta}\partial\bs{\beta}^T} &= -X^TV\left(\bs{Z}\vert M\bs{\delta},\bs{\beta}\right)X, 
\end{aligned}
\end{equation}
where $ V(\bs{Z}\vert M\bs{\delta},\bs{\beta}) $ is a diagonal matrix with elements whose values are the conditional variance of $ \bs{Z} $.

Estimation for $\bs{\theta}$ is discussed separately for the two cases. In the continuous case $ \bs{\theta} = (\sigma^2, \phi)$. For a given $ \phi $, the analytical derivatives in \eqref{eqn:updateequations} with respect to $ \sigma^2 $ are
\begin{equation} \label{eqn:dtheta}
\begin{aligned}
\frac{\partial \ln f(\bs{\delta}\vert\bs{\theta})}{\partial{\sigma^2}} &= - \frac{m}{2\sigma^2} + \frac{1}{2(\sigma^2)^2}\bs{\delta}^T\bs{\delta} \\ \frac{\partial^2\ln f(\bs{\delta}|\bs{\theta})}{\partial{(\sigma^2)^2}} &=   \frac{m}{2(\sigma^2)^2} - \frac{1}{(\sigma^2)^3} \bs{\delta}^T\bs{\delta}.
\end{aligned}
\end{equation}
The analytical derivatives with respect to $\phi$, however, are not available, as the projection matrix $ M = M_\phi $ is related to $ \phi $ in a complicated fashion. Therefore, we estimate $ \phi $ via a numerical routine. At the $ t^{th} $ iteration, we first update $ \left(\bs{\beta}(\phi)^{(t+1)},\sigma^2(\phi)^{(t+1)}\right) $ conditioning on $ \bs{\psi}^{(t)} $; they are plugged into the approximated $ Q $-function $ \hat{Q}(\bs{\psi}, \bs{\psi}^{(t)}) $ to obtain $ \hat{Q}(\phi) $. We then perform a numerical search on the neighboring values of  $ \phi^{(t)} $ to find $ \phi^{(t+1)} $ that satisfies $\hat{Q}(\phi^{(t+1)}) > \hat{Q}(\phi^{(t)})$. In the discrete case $ \bs{\theta} = \tau $. The derivatives with respect to the smoothing parameter $ \tau $ are
\begin{equation*}
\frac{\partial \ln f(\bs{\delta}|{\tau})}{\partial{\tau}} =\frac{m}{2\tau} - \frac{1}{2} \bs{\delta}^TQ_\delta\bs{\delta} \text{ and }
\frac{\partial^2\ln f(\bs{\delta}|{\tau})}{\partial{\tau^2}} = -\frac{m}{2\tau^2}.
\end{equation*}

The uncertainty of the estimates can be quantified by the asymptotic standard errors for the MLE, which is approximated using the observed information matrix $I(\bs{\psi};\bs{Z}) = -\partial^2/\partial\bs{\psi}\partial\bs{\psi}^T\ln L( \bs{\psi}; \bs{Z}) $. Often it is readily obtainable from the last iteration of the maximization step if a gradient approach is used in the M-step \citep[][Sec. 4]{EMbook},
\begin{equation}\label{eqn:se}
\begin{aligned}
I(\bs{\psi};\bs{Z}) 
&= \mathcal{I}_c(\bs{\psi};\bs{Z})- E\left[ S_c(\bs{\psi};\bs{Z},\bs{\delta})S_c^T(\bs{\psi};\bs{Z},\bs{\delta})\mid \bs{Z}\right] \\
&+ E\left[ S_c(\bs{\psi};\bs{Z},\bs{\delta}) \mid \bs{Z}\right]E\left[ S_c^T(\bs{\psi};\bs{Z},\bs{\delta})\mid \bs{Z} \right],
\end{aligned}
\end{equation}
where $\mathcal{I}_c(\bs{\psi};\bs{Z}) = -Q^{\prime\prime}$ is the conditional expectation of the complete-data information matrix, and $S_c(\bs{\psi};\bs{Z},\bs{\delta})$ is the first derivative of the conditional log complete-data likelihood. The observed information matrix only need to be evaluated once at the last EM iteration with little additional computation, as the first term is a result from the EM, the second term is approximated in the last EM iteration, and the third term is zero under the MLE. The parametric bootstrap \citep{bootstrap} is another useful approach for obtaining standard errors of the estimates. For the parametric bootstrap, we first fit the projection-based model to the data to obtain parameter estimates. Then, multiple data sets are simulated from SGLMM. For each simulated data set, we again fit the projection-based model. Finally, we estimate the standard errors from the point estimates. 

Similar to the traditional SGLMM, the projection-based models do not have a closed form expression for the conditional expectation required in the E-step. We derive two approximation methods for the projection-based model, which results in two variants of the EM algorithm.

\subsection{MCEM Algorithm}\label{Sec:MCMCEM}
We develop an automated MCEM algorithm for the projection-based models, where the conditional expectations are approximated using MCMC samples. The MC sample size at each EM iteration is selected automatically which reduces the amount of manual tuning. The E-step includes\\
\textbf{ (a) Simulation}: obtain an MCMC sample $ \bs{\delta}^{(t,1)}, \dots, \bs{\delta}^{(t,k_t)} $ with a sample size of $ k_t $, from $ f_{\delta|Z} ( \bs{\delta}|\bs{Z},\bs{\psi}^{(t)} ) $ under the current $ \bs{\psi}^{(t)} $.\\
\textbf{ (b) Monte Carlo integration}: approximate conditional expectation using average, $$
\hat{Q}(\bs{\psi}, \bs{\psi}^{(t)}) = \frac{1}{k_t}\sum_{k=1}^{k_t} \ln f_{Z,\delta}(\bs{Z},\bs{\delta}^{(t,k)}; \bs{\psi}).$$

\subsubsection{MCMC Sampling}\label{sec:Estep}
MC samples from the conditional distribution can be easily obtained using an MCMC algorithm \citep{Robert2005}. The projection-based models have reduced-dimensional and de-correlated random effects; this is advantageous in constructing MCMC over the traditional SGLMMs. We use the Metropolis-Hastings algorithm with a multivariate normal proposal function for sampling $\bs{\delta}$. 

Several strategies are utilized for constructing an efficient MCMC algorithm. (1) We use adaptive MCMC \citep{RobertsRosenthal} to avoid tedious manual tuning and to maintain desirable acceptance rate; for the $(t+1)^{th}$ EM iteration, we adjust the variance of the proposal function using $ 0.95\times 2.38^2/m\times \Sigma_t+0.05\times (0.1)^2/m\times \bs{I}_m $, where $ \Sigma_t $ is the sample covariance of the target distribution based on the current $ k_t $ sample. (2) We initiate the MCMC using the last iteration of MCMC from the previous EM update, $ \bs{\delta}^{(t+1,1)} = \bs{\delta}^{(t,k_t)}$, to obtain a good starting value. (3) We  automatically adjust the MC sample size for each EM iteration using the ascent-based approach proposed by \cite{ascentMCEM} in order to recover EM's ascent property and allocate computing resources efficiently. A sketch of the ascent-based approach and our implementation are provided below.

\subsubsection{Sample Size Selection}
The MC sample size $k_t$ at the $t^{th}$ EM iteration is chosen automatically such that it increase the $Q$-function with a high probability.  At  the t$^{th}$ iteration, we seek $ \psi^{ (t+1)} $ that maximizes the Q-function with a high probability. Let $$\triangle Q(\bs{\psi}^{(t+1, k_t)},\bs{\psi}^{(t)}) \equiv  Q(\bs{\psi}^{(t+1, k_t)},\bs{\psi}^{(t)}) -  Q(\bs{\psi}^{(t)},\bs{\psi}^{(t)})$$ be the change in the $Q$-function. Its approximation $\triangle \hat{Q}(\bs{\psi}^{(t+1, k_t)},\bs{\psi}^{(t)})$, or simply $\triangle \hat{Q}$, computed from the MC integration step, when suitably normalized, has a limiting normal distribution centered at $\triangle {Q}$ and a variance $\sigma_{\triangle Q}^2$. Let $z_\alpha$ be the $(1-\alpha) ^{th}$ percentile of a standard normal random variable $z$. We compute the asymptotic lower bound, $\triangle \hat{Q}- z_\alpha \text{ASE}$, where ASE denotes the asymptotic standard error estimated using batch means \citep{FlegalHaranJones2008}.  If the asymptotic lower bound is negative, then the increase in the $Q$-function is indistinguishable from zero due to a large MC error, indicating that a larger sample size is required. Using this as a guideline, we increase sample size from $k_t$ to $k_t + k_t/2$ until the asymptotic lower bound is positive.

In an EM algorithm, we will always choose  $\psi^{(t+1)}$ to maximize $ Q(\psi , \psi^{ (t)}) $. However, in the MCEM algorithm, $Q$ is approximated using a MC sample, therefore, it is subject to MC errors. When the increase in $ Q $ is larger than the MC error, we can be fairly certain that  $\psi^{(t+1)}$ maximizes $ Q $ without the need of increasing the sample size. This is typically the case at first few iterations of the EM algorithm. As we update the parameter estimates, the increase in $ Q $ becomes smaller as we are getting near the maximum, therefore, a larger MC sample size is needed to reduce the MC error in determining $\psi^{(t+1)}$.

Figure \ref{fig:cont_poi_est} shows that the required MC sample sizes are small in the early EM iterations, and gradually increase as the parameter estimates get near the optimal region. To ensure that a large enough MC sample is obtained at the first EM iteration to fully explore the parameter space and to estimate the correlation structure of the target distribution, we run the MCMC until the effective sample size is at least twice the dimension of the target distribution. 

\subsubsection{Approximate Conditional Expectations}
After obtaining the MCMC samples $ \bs{\delta}^{(t,k)}$, $k=1,\dots,k_t $ from $ f_{\delta|Z} ( \bs{\delta}|\bs{Z},\bs{\psi}^{(t)} ) $, the conditional expectations in \eqref{eqn:dbeta} is approximated by $ \frac{1}{k_t}\sum_k X^T(\bs{Z}-E[\bs{Z}\vert M\bs{\delta}^{(t,k)},\bs{\beta}^{(t)}]) $ and 
$ \frac{1}{k_t}\sum_k X^T V(\bs{Z}|M\bs{\delta}^{(t,k)},\bs{\beta}^{(t)})X$. The conditional expectations in \eqref{eqn:dtheta} involves computing $E[\bs{\delta}^T\bs{\delta} | \bs{Z}, \bs{\psi}^{(t)}]$ which is approximated by $ \frac{1}{k_t}\sum_k{\bs{\delta}^{(t,k)}}^T\bs{\delta}^{(t,k)} $. The numerical maximization of $\phi$ is reduced to computing the difference \begin{equation}
\begin{aligned}
  \hat{Q}(\phi^\ast) - \hat{Q}(\phi^{(t)})= & 
-\frac{1}{2}\left(\sum_{i=1}^m \ln(d_{\phi^\ast,i}) - \sum_{i=1}^m \ln (d_{\phi,i})\right)-\frac{1}{2\sigma^{2,(t+1)}}\times\\ 
& \frac{1}{k_t}\sum_{k=1}^{k_t} \left(M\bs{\delta}^{(t,k)}\right)^T\left(U_{\phi^\ast} D_{\phi^\ast}^{-1}U_{\phi^\ast}^T -U_{\phi}D_{\phi}^{-1}U_{\phi}^{T}\right)\left(M\bs{\delta}^{(t,k)}\right)\\
\end{aligned}
\end{equation} 
where $\phi^\ast$ is a neighboring value of $\phi^{(t)}$. The above comparison is performed for several neighboring values, and the one with the largest increase is set to $\phi^{(t+1)}$. The major computation involved is computing the eigencomponents of $R_{\phi^\ast}$; performing eigendecompositions several iterations for data size up to a couple of thousands is relatively fast, and we can parallelize it for multiple $\phi$ using a multicore machine. If the data size is much larger than a few thousands, we can approximate the eigencomponents using a probabilistic Nystr\"om's approximation algorithm (see supplementary materials).

\subsection{LAEM algorithm}\label{sec:LAML}
The LA is a fast alternative to MC integration for approximating the conditional expectations. It is performed for every EM iteration and includes two parts: \\
\textbf{(a) Gaussian approximation}: approximate the conditional density function $ f_{\delta|Z} \left( \bs{\delta}|\bs{Z}, \bs{\psi} \right)$ with a Gaussian distribution $f_G \left( \bs{\delta}|\bs{Z}, \bs{\psi} \right)$.\\
\textbf{(b) Taylor expansion} for functions of the random effects  $h(\bs{\delta})$ and approximate $ E_{\delta|Z} \left[ h(\bs{\delta})\vert \bs{Z},\bs{\psi} \right] $ with $ E_G\left[ \tilde{h}(\bs{\delta})\vert \bs{Z},\bs{\psi} \right] $, where $\tilde{h}(\bs{\delta})$ denotes the approximation to ${h}(\bs{\delta})$ and the expectation is taken with respect to $ f_G $.

\subsubsection{Gaussian Approximation}\label{sec:GaussApprox}
For the projection-based model, the conditional density function has the form
\begin{equation*}
f_{\delta|Z} \left( \bs{\delta}|\bs{Z}, \bs{\psi}\right) \propto  \exp\left\lbrace -\frac{1}{2} \bs{\delta}^T Q_{\delta}\bs{\delta} + \sum_i \ln f_{Z_i|M\bs{\delta}} \left( Z_i|M\bs{\delta},\bs{\beta} \right)  \right\rbrace. 
\end{equation*} 
We approximate it with a Gaussian distribution whose mean is matched with the mode and variance with the inverse of the negative Hessian of $f_{\delta|Z} \left( \bs{\delta}|\bs{Z}, \bs{\psi}\right)$ evaluated at the mode. 

We first Taylor expand $ \sum_i \ln f_{Z_i|M\bs{\delta}} \left( Z_i|M\bs{\delta},\bs{\beta} \right) $ to the second order around an initial guess $ \bs{\delta}^{(0)} $. This will give a quadratic form in $ \bs{\delta} $,  for example, for count observations this becomes $-\frac{1}{2} \bs{\delta}^TM^TD_2M\bs{\delta} + \bs{\delta}^TM^T(Z-d_1 + D_2H\bs{\delta}^{(0)}) + \textit{const},$
where $ D_2 = \texttt{diag}(\exp(X\bs{\beta} + M\bs{\delta}))\mid_{\bs{\delta} = \bs{\delta}^{(0)}}$ is an $ n\times n $ diagonal matrix, $ d_1 = \exp(X\bs{\beta} + M\bs{\delta}) \mid_{\bs{\delta} = \bs{\delta}^{(0)}}$ is an $n$-dimensional vector and $\textit{const}$ is a constant that does not depend on $\bs{\delta}$. For Poisson observation model, $D_2$ and $d_1$ have the same elements, but this is not always the case for the exponential family; as an example, Gaussian approximation for the binary case is shown in the supplementary materials. 

The conditional density function is then approximately $$f_{\delta|Z} \left( \bs{\delta}|\bs{Z}, \bs{\psi}\right) \approx \exp\left\lbrace -\frac{1}{2} \bs{\delta}^T Q\bs{\delta}  + \bs{\delta}^TM^T\bs{b}  \right\rbrace,$$ where $ Q = M^TD_2M + Q_{\delta} $, $ \bs{b} =   M^T\left(Z-d_1+D_2M\bs{\delta}^{(0)}\right) $.  The approximation has a form similar to the density function of a multivariate Normal $N\left(Q^{-1}b, Q^{-1}\right)$. We then find the mode $ \bs{\delta}^{\ast} $ using Newton-Raphson by solving $ \bs{\delta} = Q^{-1}b $ iteratively until convergence. Once obtaining the mode, the mean and variance of the Gaussian approximation $ f_G\left( \bs{\delta}|\bs{Z}, \bs{\psi}\right) $ are $E(\bs{\delta}\mid \bs{Z},\bs{\psi}) = \bs{\delta}^{\ast}$ and $V(\bs{\delta}\mid \bs{Z},\bs{\psi}) = Q^{-1}\mid_{\bs{\delta}=\bs{\delta}^\ast}$, respectively.

\subsubsection{Approximate Conditional Expectations} The terms to be approximated in the conditional expectations has the form \newline $E_{\delta|Z}\left[h(X_i\bs{\beta}+M_i\bs{\delta})\mid\bs{Z_i},\bs{\psi}^{(t)}\right]$. We use $\tilde{h}(X_i\bs{\beta}+M_i\bs{\delta})$ to denote the second order Taylor expansion of $ h(X_i\bs{\beta}+M_i\bs{\delta}) $ around $ \bs{\delta}^\ast $, then
\begin{equation}
\begin{aligned}
 \tilde{h}(X_i\bs{\beta}  +M_i\bs{\delta})  =  & h(X_i\bs{\beta} +M_i\bs{\delta}^\ast)   + (\bs{\delta}-\bs{\delta}^\ast)^T \left( h^\prime(X_i\bs{\beta}+M_i\bs{\delta}^\ast)\times M_i^T\right) \\
& + \frac{1}{2}  (\bs{\delta}-\bs{\delta}^\ast)^T \left( h^{\prime\prime}(X_i\bs{\beta}+M_i\bs{\delta}^\ast)\times M_i^T M_i \right) (\bs{\delta}-\bs{\delta}^\ast),
\end{aligned}
\end{equation}
where $ M_i $ is the $ i^{th} $ row of the projection matrix M, and $h^\prime(x^\ast) = dh(x)/dx\mid_{x=x^\ast}$. We take the expectation of the above with respect to $ f_G(\bs{\delta}\mid\bs{Z},\bs{\psi}^{(t)}) $ and obtain the following,
\begin{equation}
\begin{aligned}
 E_G\left[\tilde{h}(X_i\bs{\beta}+M_i\bs{\delta})\mid\bs{Z_i},\bs{\psi}^{(t)}\right] 
&= h(X_i\bs{\beta}+M_i\bs{\delta}^\ast)  +  E\left[ (\bs{\delta}-\bs{\delta}^\ast)^T|\bs{Z},\bs{\psi}^{(t)}\right] \left( h^\prime(X_i\bs{\beta}+M_i\bs{\delta}^\ast)M_i^T\right) \\
& + \frac{1}{2}  tr\left\lbrace E \left[ (\bs{\delta}-\bs{\delta}^\ast)(\bs{\delta}-\bs{\delta}^\ast)^T |\bs{Z},\bs{\psi}^{(t)}\right] \left( h^{\prime\prime}(X_i\bs{\beta}+M_i\bs{\delta}^\ast)M_i^T M_i \right)\right\rbrace\\
&= h(X_i\bs{\beta}+M_i\bs{\delta}^\ast) + \frac{1}{2}  tr\left\lbrace Q^{-1} \left( h^{\prime\prime}(X_i\bs{\beta}+M_i\bs{\delta}^\ast) M_i^T M_i \right)\right\rbrace,\\
&= h(X_i\bs{\beta}+M_i\bs{\delta}^\ast) + \frac{1}{2}  \left( h^{\prime\prime}(X_i\bs{\beta}+M_i\bs{\delta}^\ast) M_i Q^{-1} M_i^T \right).
\end{aligned}
\end{equation}
The second equality holds as $ E\left[(\bs{\delta}-\bs{\delta}^\ast)^T\vert\bs{Z},\bs{\psi}^{(t)}\right]=0 $   and
$ E\left[ (\bs{\delta}-\bs{\delta}^\ast)(\bs{\delta}-\bs{\delta}^\ast)^T \vert\bs{Z},\bs{\psi}^{(t)}\right] = Q^{-1}\mid_{\bs{\delta}=\bs{\delta}^\ast}$. 

\subsection{Rank Selection}\label{sec:rkselection}
The projection-based model is based on spatial filtering \citep{griffith2013spatial} and principal component analysis. We can fit non-spatial generalized linear models with predictors $ X $ and synthetic spatial variables $ U_mD_m^{1/2} $ for $ m = 1,2,\dots $ where the eigencomponents are computed from $ R_{\phi^{(0)}} $ using an initial range value $\phi^{(0)}$. Then the rank can be selected based on variable selection criterion such as AIC. This serves as a general guideline for selecting the initial rank. Based on this, we can then fit a few models with different ranks and perform cross-validation or use penalized-likelihood criteria such as AIC or BIC to determine the final model.

\subsection{Computational Benefit from the Projection-Based Models}
Both the MCEM and LAEM algorithms can be used for inference for the traditional SGLMM. However, several computational challenges make it prohibitive when the data size is large. In the MCEM algorithm, (1) MC sampling from $ f_{W|Z} ( \bs{W}|\bs{Z},\bs{\psi}^{(t)} ) $ requires manipulating a large $ n\times n $ matrix, which is computationally slow even for high performance computers. (2) The random effects are also highly correlated, making it difficult to construct an efficient sampling algorithm. The reparameterization in the  projection-based model replaces the original random effects $\bs{W}$  with a  much smaller number of new random effects $\bs{\delta}$ that are also de-correlated, resolving the above two challenges simultaneously. Using the LAEM algorithm to fit the traditional
SGLMM requires Gaussian approximation to $ f_{{W}|Z} $, which has the same dimension of the observations. For a relatively small number of data points, say hundreds, the LA is fast, but as the data size grows, this will become computationally challenging. The projection-based model provides a viable solution as the dimension of random effects is reduced significantly compared to the original data size. The Gaussian approximation $ f_G\left( \bs{\delta}|\bs{Z},\bs{\psi}\right) $ now has the same dimension as the chosen rank, which is typically much less than a hundred, therefore matrix manipulation involving its covariance matrix $Q^{-1}$ is fast.

\subsection{Spatial Prediction}\label{sec:pred}
SGLMMs are also often used for interpolation/prediction at unsampled locations. We describe interpolation using the projection-based models with a focus on the continuous case, since it is often less of interest for the discrete case in practice. Let $\mathcal{S}^\ast = \left\lbrace \bs{s}^\ast_1,\dots,\bs{s}^\ast_{n^\ast}\right\rbrace $ be a set of unsampled locations. In the projection-based model, the covariance between $\bs{W}^\ast$ at $\mathcal{S}^\ast$ and the reparameterized random effects $\widetilde{\bs{W}}=U_\phi D_\phi^{1/2}\bs{\delta}$ is 

\begin{equation*}
\text{cov}\left\lbrace\left(
\begin{matrix}
\widetilde{\bs{W}}\\
\bs{W}^\ast
\end{matrix}\right)\right\rbrace = \left(\begin{matrix}
\left(U_\phi D_\phi^{-1/2}\right)^T\Sigma_{\bs{\theta}} \left(U_\phi D_\phi^{-1/2}\right) & \left(U_\phi D_\phi^{-1/2}\right)^T\Sigma_{\bs{\theta},s\ast}\\
\Sigma_{\bs{\theta},\ast s} \left(U_\phi D_\phi^{-1/2}\right) &\Sigma_{\bs{\theta},\ast\ast}
\end{matrix}\right), 
\end{equation*}
where $\left(U_\phi D_\phi^{-1/2}\right)^T\Sigma_{\bs{\theta}} \left(U_\phi D_\phi^{-1/2}\right)$ is simply $\sigma^2I_{m\times m}$.
The best linear unbiased predictor (BLUP) of $\bs{W}^\ast$ given $\widetilde{\bs{W}}$ is therefore  $\bs{W}^\ast|\widetilde{\bs{W}},\bs{\theta}\sim$ $ \text{MVN}(\mu_{\bs{W}^\ast|\widetilde{\bs{W}}}, \Sigma_{\bs{W}^\ast|\widetilde{\bs{W}}})$ \citep{Stein1999}, where 
$
\mu_{\bs{W}^\ast\mid \widetilde{\bs{W}}} = \frac{1}{\sigma^2}\Sigma_{\bs{\theta},\ast s} (U_\phi D_\phi^{-1/2})\widetilde{\bs{W}}$ and $
\Sigma_{\bs{W}^\ast\mid \widetilde{\bs{W}}} = \Sigma_{\bs{\theta},\ast\ast}-\frac{1}{\sigma^2}\Sigma_{\bs{\theta},\ast s} (U_\phi D_\phi^{-1}U_\phi^T)\Sigma_{\bs{\theta},s\ast}.
$
If MCEM algorithm is used, to make spatial prediction we will draw from the above multivariate normal distribution for each MCMC sample of the random effects $\widetilde{\bs{W}}=U_\phi D_\phi^{1/2}\bs{\delta}^{(k)}, k=1,\dots,k_t$. If the LAEM is used, then $\widetilde{\bs{W}}$ is approximated by $U_\phi D_\phi^{1/2}\bs{\delta}^{\ast}$, where $\bs{\delta}^{\ast}$ is the mode from the Gaussian approximation at the last EM iteration. 

The prediction uncertainty from the MCEM is typically larger than the LAEM, as it incorporates the random effect uncertainty in prediction while the LAEM does not. However, both methods do not account for regression parameter uncertainty. If the research goal is parameter estimation or as a quick data exploratory tool, both algorithms are appropriate. If assessing spatial prediction uncertainty is the main focus, then one should keep in mind that the uncertainty from these algorithms is smaller, and therefore a fully Bayesian approach \citep[cf.][]{GuanHaran2016}  would be more appropriate.  

\section{Simulation Study}\label{Sec:simulation}
We study the proposed algorithms for both spatial counts and binary observations and for both continuous and discrete spatial domains. We present the results for the count data in a continuous spatial domain below. Results for count data on a lattice and binary data are similar and therefore presented in the supplementary materials.

\subsection{Count Data in a Continuous Spatial Domain}\label{sec:simu_cont}
We simulate $n$=1400 random effects $ \bs{W} $ in the unit domain $ [0,1]^2 $ from the Mat\'ern covariance model. Conditional on $ \bs{W} $, we simulate $Z_i$ from $\text{Poisson}(\mu_i)$ with $\log(\mu_i)$=$x_{i,1} +x_{i,2} + W_i$, where $x_{i,1},x_{i,2}$ are the xy-coordinates of $W_i$. 
The data are generated using $\bs{\beta} = (1,1)^T $, $ \nu=1.5 $, $\sigma^2 =1$ and $\phi=0.073$ and 0.18. The range values correspond to effective range (defined as the distance at which the correlation is 0.05) $r$ = 0.2 and 0.5, respectively. The first 1,000 observations are located randomly in the spatial domain and are used for model fitting, while the rest are located on a $20\times20$ grid and are used for testing. 

We suggest obtaining initial value of the regression coefficient and residual variance from fitting a non-spatial generalized linear model (GLM). It is typically difficult to obtain an estimate for the range parameter $\phi$ from the non-Gaussian observations; therefore, we take roughly half of the spatial domain as the initial value. 


We first fit the projection-based model for two simulated data sets, each of which is simulated with different values of effective range, to investigate the parameter estimates and prediction performance of different ranks. Based on the proposed initial rank selection, the required ranks are 90 and 50 for $r$ = 0.2 and 0.5, respectively. Then we fit both algorithms using a few different ranks near the initial selection, for example, ranks 70 to 110 with an increment of 10 for the first case. 

Results for $r$=0.2 are presented here because the conclusion for $r$=0.5 is similar. The initial values estimated from GLM are $\bs{\beta}^{(0)} = (0.78,    1.94)^T$ and $\sigma^{2{(0)}} = 2.22$. For MCEM, we have used $\alpha$=0.15, $\gamma$=0.05, and $\epsilon$=0.01 (the latter two are tuning parameters in the MCEM to determine the stopping criteria; see supplementary materials S.3).  {Parameter estimates and prediction performance are summarized in Table \ref{tab:count_cont_rank}. The results suggest that rank 90 or 100 seems to be sufficient, as the parameter estimates become stable and the mean squared prediction error (MSPE) improvement decreases. The variability in the MSPE from MCEM algorithm can be a result from varying MCMC sample size and therefore providing different accuracy. The variability in the MSPE from LAEM algorithm can be a result of the LA. We conduct a simulation study with 50 replicates to assess how MSPE varies as rank increases. Table \ref{tab:count_cont_rank_rep} reports the median and the 95\% quantile for the MSPE and computational time from a simulation study with 50 replicates.  We can see a more clear overall decreasing trend in MSPE as the rank increases, however the variability in MSPE is also large for each rank. Here, the MSPE variability represents a combined variability of data sampling as well as fitting the proposed algorithms due to MCMC sampling and the LA.}

We also notice that the prediction from the LAEM algorithm consistently under-performs compared to the MCEM. This suggests that the random effects estimated from LA are not as well as the MC approach. Figure \ref{fig:cont_poi_re} shows the predicted linear component in the conditional mean from the two algorithms. The computational time  using the MCEM algorithm is typically around 2-3 minutes and less than 1 minute for the LAEM (Table \ref{tab:count_cont_rank_rep}). These are much faster than the fully Bayesian with MCMC approach proposed in \citep{GuanHaran2016}, which took roughly 4 hours for the same data size. 
\begin{center}
	
 [Table 1 here] 
 
 [Table 2 here] 
 
 [Figure 1 here] 
 \end{center}

To monitor convergence and study the robustness of the two algorithms to initial value, we run both algorithms for a fixed number of iterations at three different starting values. We found that the LAEM algorithm is more sensitive to initial value than MCEM. For the same initial values tested, the MCEM typically converges within 40 EM iterations, while the LAEM may diverge if the initial value is not carefully selected. For the simulated data, the initial value obtained from GLM works well for both algorithms.

Since MCEM is more robust to different initial values, here we focus on illustrating the performance of MCEM. Figure \ref{fig:cont_poi_est}(a,b) shows the parameter estimates at each iteration from the MCEM algorithm; the parameter estimates converge to the same values. Figure \ref{fig:cont_poi_est}(c) shows the MC sample sizes at each EM iteration; most of the simulation efforts are spent in the first and the last 2-3 EM iterations. Typically, when the stopping threshold is reached (indicated by the vertical dashed line), the ascent-based MCEM algorithm provides a large MC sample. This is a desirable feature, since the last MC sample is used in subsequent analyses, for instance, for estimating the observed information matrix and spatial prediction. Finally, the integrated log-likelihood function corresponding to different starting values stabilizes as the EM iteration increases.

\begin{center}
	[Figure 2 here] 
\end{center}

We conduct a simulation study with 100 replicates to study the distribution of the point estimates. Figure \ref{fig:poisson_cont_box} shows the boxplots of the estimates; it appears that for both algorithms $\hat{\bs{\beta}}$ are unbiased, while $\hat{\bs{\theta}}$ have positive biases. 

\begin{center}
	[Figure 3 here]
\end{center}

We compare the interval estimation based on the observed information matrix and bootstrap. For the latter, a bootstrap sample of 100 replicates was used to compute the confidence intervals for each simulated data set.
The coverages based on the observed information matrix are around 15\%, much lower than the nominal rate 95\%, whereas the coverages based on bootstrap are near 95\%, because the confidence intervals (CIs) provided by the observed information matrix is much narrower than the ones from bootstrap and therefore missed the true values.

%

\subsection{Method Comparison}
We compare MCEM and LAEM algorithms without dimension reduction to a Bayesian approach (\textit{spBayes}) for fitting the SGLMM. Two scenarios are considered to assess how the methods perform with and without spatial confounding. For the spatial-confounding case, the simulation setup is similar to Section 5.1 where covariates are the xy-coordinates of the observations. For the no-spatial-confounding case, covariates are independent and identically distributed standard normal random variables. Data with size n=300 are generated. For the predictive process approach we run the MCMC algorithm for 50,000 iterations, discarding the first 25\% for burn-in. 

Table \ref{tab:method_comp}(a) shows the parameter estimates under the no-spatial-confounding case. The results from the reduced-dimensional model are similar to the ones without dimension reduction when the same algorithm is used. This is expected since the rank selected by the rank selection guidelines provided in Section 4.4 is sufficient in capturing most of the spatial variability of the spatial random effect. This is also confirmed in Figure \ref{fig:method_comp_reest_noconfound}, which plots the estimated random effects against the true values for the MCEM with rank 90 and full rank, as well as the predictive process approach. We summarize these comparisons in scatter plots. The plots are very similar for the LAEM algorithm comparisons so we have not included them. 
\begin{center}
	[Table 3 here]
	
	[Figure 4 here]
\end{center}
Table \ref{tab:method_comp}(b) shows the parameter estimates under the spatial-confounding case. The results from the MCEM and LAEM algorithms are similar as they both impose orthogonality between the random effects with the fixed effects. The regression parameters are both smaller than the true values and the 95\% confidence intervals tend to miss the truth. The inference results observed here are consistent with \cite{ephraim2015} which notes that RSR may elevate the Type 1 error rate when spatial confounding is present. We recommend obtaining confidence intervals using parametric bootstrap, as it provides near nominal coverage rate as described in the last paragraph of Section 5.1. A major advantage of using the RSR is in estimating the random effects. Figure \ref{fig:method_comp_reest_confound} shows the random effect estimates obtained from the MCEM algorithms outperforms the predictive process approach. The underperformance for the predictive process approach is due to the challenges in MCMC sampling when the posterior distribution is high-dimensional and highly correlated. 
\begin{center}
	[Figure 5 here]
\end{center}
	
\subsection{Large Data Example}\label{sec:simu_cont_large}
To study how long our method takes in the context of a large data set, we fitted MCEM and LAEM with rank 100 to a simulated data set with size $ n = 50,000 $. The data are generated from the no-spatial-confounding scenario so the covariates are independent and identically distributed standard normal random variables. Our estimate for $(\beta_1, \beta_2, \tau, \phi)$ is (1.07,1.06,1.03,0.06) from LAEM and (1.01,1.01,1.10,0.07) from MCEM. We see that the coefficient estimates from the MCEM algorithm are closer to the true value $(\beta_1, \beta_2, \tau, \phi)=(1,1,1,0.073)$ than LAEM and the pattern of the estimated spatial random effects from MCEM is also closer to the true pattern (Figure \ref{fig:massive}). As in the simulation study, here too MCEM is apparently preferable to LAEM in terms of accuracy, though it is more computationally expensive at 10.84 hours versus 1.67 hours for the LAEM.
\begin{center}
	[Figure 6 here]
\end{center}

\section{Data Analysis}\label{Sec:application}

\subsection{US Infant Mortality Count Data}
We fit the projection-based model for the county-level US infant mortality from 2002 to 2004, a data set analyzed in \cite{HughesHaran} under a Bayesian approach. The response variable is the 3-year average number of infant deaths before the first birthday, and the predictors are the rate of low birth weight (low), the percentage of black residents (black), the percentage of Hispanic residents (Hisp), a measure of income inequality (the Gini coefficient proposed by Gini, 1921), a composite score of social affluence (aff, proposed by Yang et al., 2009) and residential stability (stab, an average z-score of two variables). Similar to \cite{HughesHaran}, we use the 3-year average number of live births as an offset to adjust for the population difference in these counties. Our results are comparable to the ones from the Bayesian inference with MCMC in \cite{HughesHaran}. Therefore, the point and interval estimates are summarized in the supplementary materials. 


\subsection{Forest/Non-forest Land Type Data}
The land type of a region, whether it is covered by forest or non-forest, is often of interest for economic and environmental reasons. Spatial regression can be used for assessing the relationship between forest/nonforest binary response and potential covariates while accounting for the residual spatial dependence. We use a data set analyzed in \cite{Berrett2016}. The response variable is 2005 Land Cover Type Yearly Level 3 Global 500 m (MOD12Q1 and MCD12Q1) data from Moderate Resolution Imaging Spectroradiometer (MODIS); the MODIS land cover data are categorized into two types, forest and nonforest. The study region is a 24$\times$24 regular grid between 17$^{\circ}$-19$^{\circ}$ N and 98$^{\circ}$-100$^{\circ}$ E, covering a portion of northwestern Thailand and a small part of Myanmar. We randomly sample 450 out of 576 grid cells for training, and test on the remaining for model validation. In our analysis, the observations are modeled using a Gaussian random field latent process with coordinates taken to be the centroid of the grid cells. The covariates considered are elevation, distance to the coast, distance to nearest big city, and distance to the nearest major road. 

We fit the projection-based model using the MCEM algorithm using rank 70. Based on our analysis, it appears that higher elevation, longer distance to the coast and road are associated with higher forest coverage for this area. The point and interval estimates are summarized in the supplementary materials. 


\section{Summary}\label{chap3:sec7}
We have proposed two variants of the EM algorithm that allow us to carry out maximum likelihood inference for SGLMMs. These algorithms take advantage of recent developments in dimension reduction of latent variables using projection methods \citep{HughesHaran,GuanHaran2016}. Our algorithms are computationally efficient and allow us to do maximum likelihood inference for problems where it was previously computationally prohibitive. While our goal is to do maximum likelihood inference, we also find that the algorithms are faster than corresponding MCMC-based Bayesian inference procedures in the continuous domain setting, and are comparable in speed in the discrete domain setting. Parameter estimates seem to converge quickly for both algorithms, however LAEM is less robust to initial values and may fail when initial values are far from the MLE. We recommend using initial values estimated from a fitted GLM, which worked well for both algorithms in our simulation study. An \textit{R} package to fit the proposed algorithms can be downloaded from   https://github.com/yawenguan/projSGLMM. 

Based on the simulation study results in Tables \ref{tab:count_cont_rank} and \ref{tab:count_cont_rank_rep}, where we compare performance of fitting different ranks, as well as the simulation examples for small n=300 (Table \ref{tab:method_comp}) and large n=50,000 (Section 5.3) data size. It appears that $ m  $ need not increase linearly with $ n $ and no bias is observed for increasing data size. Determining how $m$ increases with $n$ is a very complicated and context-specific question -- it depends on how many eigencomponents in our spectral decomposition are adequate for capturing the latent spatial dependence in the SGLMM. In practice we find that $m$ is often much smaller than $n$. The size of $m$ depends on the smoothness of the spatial random effects: the smoother the process, the smaller $m$ needs to be. 

The computational cost of our approach is of order $mn^2$ while the computational costs of the original approach is  of order $n^3$ per iteration of the MCMC algorithm. Hence, as $n$ gets larger, if $m$ indeed increases linearly with $n$, our algorithm will only be slightly faster than the non dimension-reduced approach. However, in a practical sense, we find that $m$ tends to be much smaller than $n$. Our  simulation study results also show that a small rank $ m=90 $ works well for data size of $ n=300, \text{or }1,000$, and for $n=50,000$ we use $m=100$. 

Maximum likelihood inference has not been as popular as Bayesian inference for SGLMMs, at least in part because of computational issues. We hope that the methodology we develop here, which addresses inference for a large class of models, including both latent Gaussian process and Gaussian Markov random field models, will allow researchers to routinely fit SGLMMs using maximum likelihood inference. We do not believe that this will entirely replace Bayesian approaches as Bayesian models allow for a greater range of flexibility in terms of adding additional hierarchies, handling missing data, and combining information from multiple variables routinely. However, for a wide range of problems, the class of SGLMMs for which we have developed a computationally efficient set of methods here, maximum likelihood inference may now be a convenient and viable option. Furthermore, because the work on practical EM algorithms for SGLMMs and other models with high-dimensional dependent latent variables has been somewhat limited, we hope that our work suggests directions for future work on scalable EM-type algorithms for such problems.

%

\bigskip
\noindent{\bf Supplementary Materials}
Contain eigencomponent approximation, Laplace approximation for binary data, stopping criterion, simulation study results for count data on a lattice and binary data, and tables for data applications.

\bigskip
\noindent{\bf Acknowledgements}
	We thank Dr. Candace Berrett for sharing the land type data.

%
%

\spacingset{1.5}
\bibliographystyle{spbasic}      
\bibliography{Reference}   
 
\afterpage{%
\begin{table} 
	\footnotesize
	\centering
	\caption{Parameter estimates and MSPE using different ranks for counts on a continuous spatial domain. Data are simulated from the Mat\'{e}rn covariance model with $\nu =1.5$, $ \sigma_2=1 $ and effective range of 0.2. The model fitting time are in minutes (min).}
	\label{tab:count_cont_rank}
	\begin{tabular}{ ccccccccccc }		
		\hline\noalign{\smallskip}
		& \multicolumn{5}{c}{LAEM} & \multicolumn{5}{c}{MCEM} \\
		Rank & $\beta_1$=1 & $\beta_2$=1 & $ \tau/\phi $=13.7 & MSPE & Time (min) & $\beta_1$=1 & $\beta_2$=1 & $ \tau/\phi $=13.7 & MSPE & Time (min)\\ 
		\noalign{\smallskip}\hline\noalign{\smallskip}
		70 & 0.35 & 1.56 & 15.96 & 4.67 & 0.30 & 0.30 & 1.55 & 16.73 & 5.19 & 1.73 \\ 
		80 & 0.36 & 1.55 & 14.97 & 3.67 & 0.31 & 0.28 & 1.54 & 17.12 & 4.04 & 8.03 \\ 
		90 & 0.37 & 1.53 & 14.22 & 3.91 & 0.35 & 0.30 & 1.51 & 15.87 & 4.58 & 3.49 \\ 
		100 & 0.33 & 1.53 & 18.81 & 3.01 & 0.52 & 0.30 & 1.53 & 22.44 & 3.23 & 2.75 \\ 
		110 & 0.36 & 1.52 & 14.77 & 3.38 & 0.48 & 0.32 & 1.52 & 14.64 & 3.29 & 2.77 \\ 
		\noalign{\smallskip}\hline
	\end{tabular}
\end{table}
\clearpage
}

\afterpage{%
\begin{table} 
	\footnotesize
	\centering
	\caption{The median and 95\% quantile of MSPE and computational time from a simulation study with 50 replicates. The data are simulated from Mat\'{e}rn covariance model with $\nu =1.5$, $ \sigma_2=1 $ and effective range of 0.2. The model fitting time are in minutes (min).}		
	\label{tab:count_cont_rank_rep}			
	\begin{tabular}{ cllll}		
		\hline\noalign{\smallskip}
		& \multicolumn{2}{c}{LAEM} & \multicolumn{2}{c}{MCEM} \\
		Rank  & MSPE & time (min) & MSPE & time (min)\\ 
		\noalign{\smallskip}\hline\noalign{\smallskip}	
		\hline
		70 & 6.1(3.1,22.9) & 0.5(0.3,2.1) & 4.8(2.2,23) & 2.1(0.9,12.5) \\ 
		80 & 6.1(2.6,24.1) & 0.5(0.3,2) & 4.8(2.3,15.9) & 1.9(0.8,8.1) \\ 
		90 & 4.7(2.5,18.1) & 0.5(0.3,1.9) & 4.3(2,19.9) & 2.3(0.9,12) \\ 
		100 & 5.2(2.4,12.7) & 0.5(0.3,1.6) & 3.7(2.3,9.1) & 2.5(0.8,12.9) \\ 
		110 & 5.4(2.3,17) & 0.5(0.3,1.6) & 3.6(1.8,10.1) & 2.1(1.1,19) \\ 
		\hline
		\noalign{\smallskip}\hline
	\end{tabular}
\end{table}
\clearpage
}

\afterpage{%
\begin{table} 
	\footnotesize
	\centering
	\caption{Parameter estimates under two cases. Data with size $ n= 300 $ are simulated from the Mat\'{e}rn covariance model with $\nu =1.5$, $ \sigma_2=1 $ and effective range of 0.2.}
	\label{tab:method_comp}
	\subfloat[no-spatial-confounding case]{
		\begin{tabular}{ cccc }		
			\hline\noalign{\smallskip}
			Method & Rank & $\beta_1$=1 & $\beta_2$=1 \\ 
			\noalign{\smallskip}\hline\noalign{\smallskip}
			LAEM & 90 & 0.95(0.88,1.02) & 1.07(1.02,1.12) \\ 
			MCEM & 90 & 1.02(0.89,1.15) & 1.13(1.03,1.24) \\ 
			LAEM & full & 0.92(0.85,1.00) & 1.04(0.99,1.09) \\ 
			MCEM & full & 1.01(0.84,1.19) & 1.12(0.98,1.25)  \\ 
			spBayes & full & 1.03(0.87,1.18) & 1.11(0.98, 1.25)\\ 
			\noalign{\smallskip}\hline
	\end{tabular}}
	\subfloat[spatial-confounding case]{
		\begin{tabular}{ cccc }		
			\hline\noalign{\smallskip}
			Method & Rank & $\beta_1$=1 & $\beta_2$=1 \\ 
			\noalign{\smallskip}\hline\noalign{\smallskip}
			LAEM & 90 & 0.91(0.75, 1.07) & 0.74(0.59,0.90) \\
			MCEM & 90 & 0.92(0.71, 1.12) & 0.76(0.55,0.96)\\ 
			LAEM & full & 0.90(0.74,1.06) & 0.76(0.61,0.91)\\ 
			MCEM & full & 0.88(0.67, 1.09) & 0.73(0.52,0.93)\\ 
			spBayes &full & 1.19(1.03, 1.34) & 1.13(0.98,  1.28) \\ 
			\noalign{\smallskip}\hline
	\end{tabular}}
	
\end{table}
\clearpage
}

\afterpage{%
\begin{figure} 
	\centering
	\subfloat{\includegraphics[width = \textwidth]{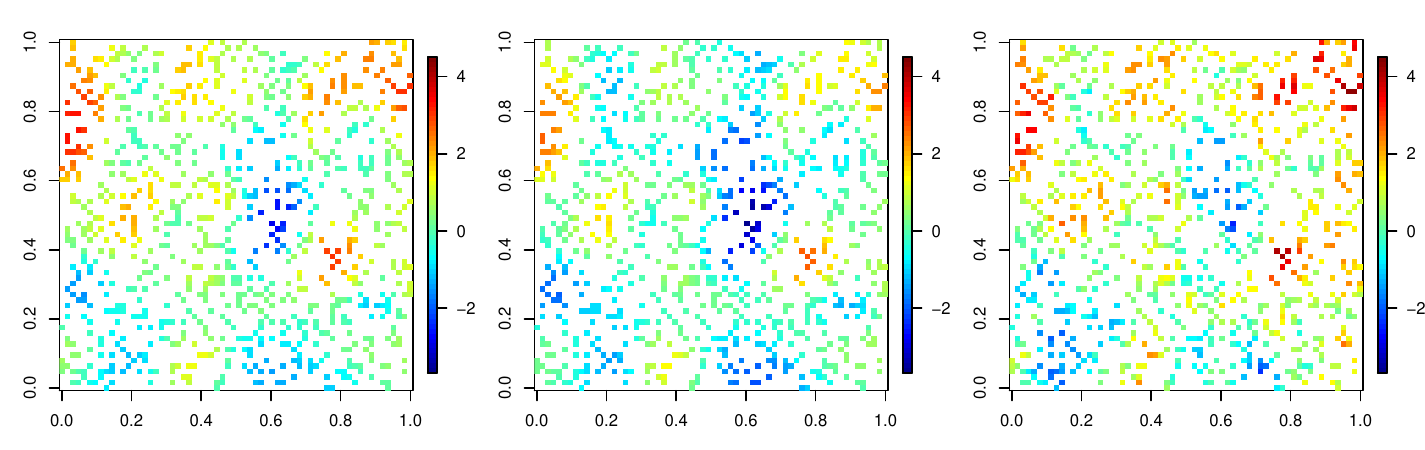}}
	\caption{Estimated linear component in the conditional mean from the LAEM (left), MCEM (center) and the true (right). Data are simulated from the Mat\'{e}rn covariance model with $\nu =1.5$, $ \sigma_2=1 $ and effective range of 0.2.}
	\label{fig:cont_poi_re}
\end{figure}
\clearpage
}
\afterpage{%
\begin{figure}
	\centering
	
	\subfloat[]{\includegraphics[width=0.4\textwidth]{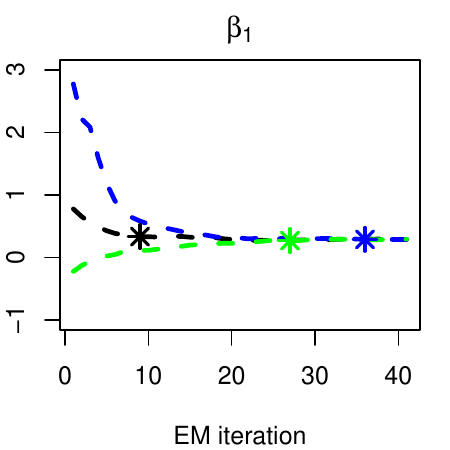}}
	\subfloat[]{\includegraphics[width=0.4\textwidth]{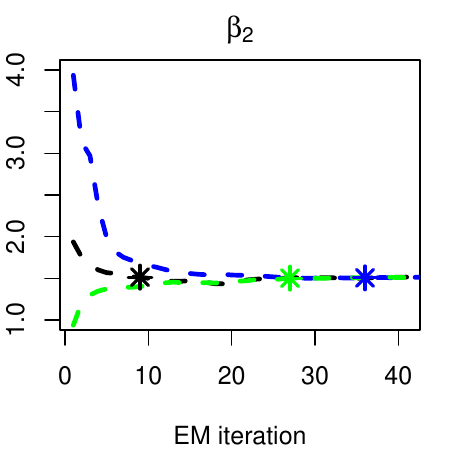}}
	\hspace{0mm}
	
	\subfloat[]{\includegraphics[width=0.4\textwidth]{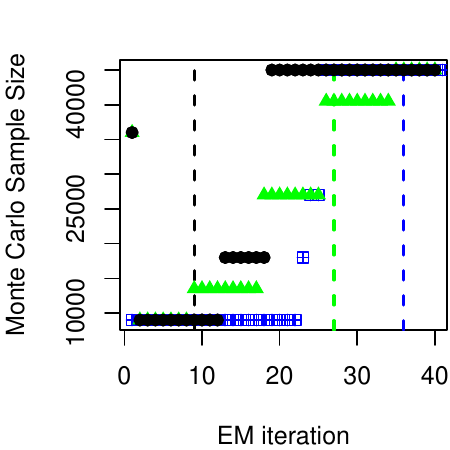}}
	\subfloat[]{\includegraphics[width=0.4\textwidth]{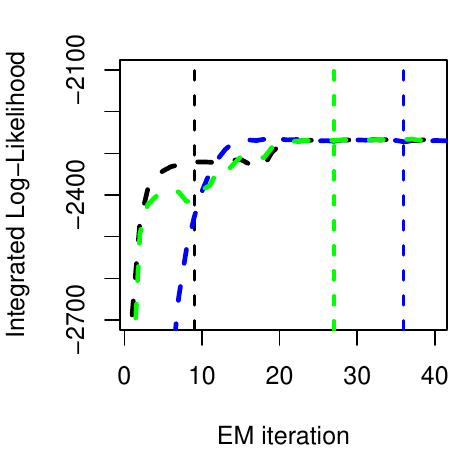}}
	\caption{Parameter estimates from MCEM with different starting values (a-b). Each color corresponds to results from the same initial values. Asterisk indicates the stopping threshold is reached. MC sample size (c) is adjusted automatically, with increasing simulation efforts as EM iteration increases. We typically obtain a large MC sample when the algorithm stops; vertical line indicates the stopping threshold is reached. Integrated log-likelihood function (d) corresponding to different starting values stabilize as EM iteration increases.}
	\label{fig:cont_poi_est}
\end{figure}
\clearpage
}

\afterpage{%
\begin{figure}
	\centering
	\subfloat{\includegraphics[width=0.9\textwidth]{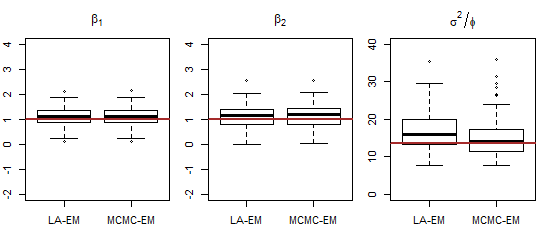}}
	
	\subfloat{\includegraphics[width=0.9\textwidth]{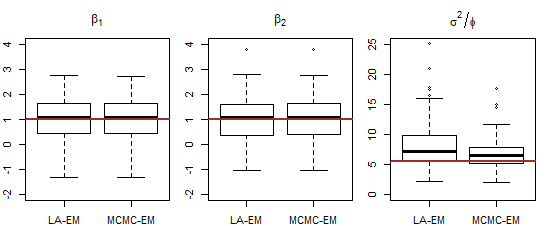}}
	\caption{Distributions of the estimates from both the LAEM and MCEM algorithm for spatial counts in continuous spatial domain. Data are simulated from the Mat\'{e}rn covariance model with $\nu =1.5$, $ \sigma_2=1 $ and effective range of 0.2 in the top panel or effective range of 0.5 in the bottom panel. It seems that $\hat{\bs{\beta}}$ are unbiased while $\hat{\bs{\theta}}$ has positive biases.}
	\label{fig:poisson_cont_box}
\end{figure}
\clearpage
}

\afterpage{%
\begin{figure}
	\centering
	\includegraphics[width = \linewidth]{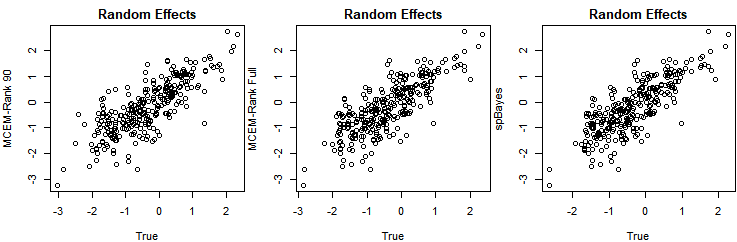}
	\caption{Estimated random effects under the no-spatial-confounding case.}
	\label{fig:method_comp_reest_noconfound}
\end{figure}
\clearpage
}

\afterpage{%
\begin{figure}
	\centering
	\subfloat{\includegraphics[width = \linewidth]{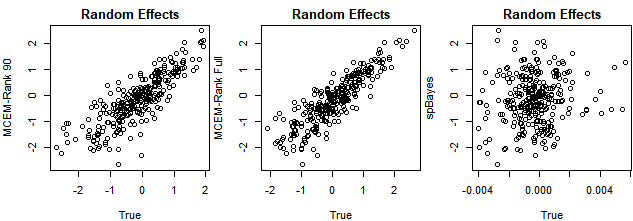}}
	
	\subfloat{\includegraphics[width = 0.4\linewidth,page=1]{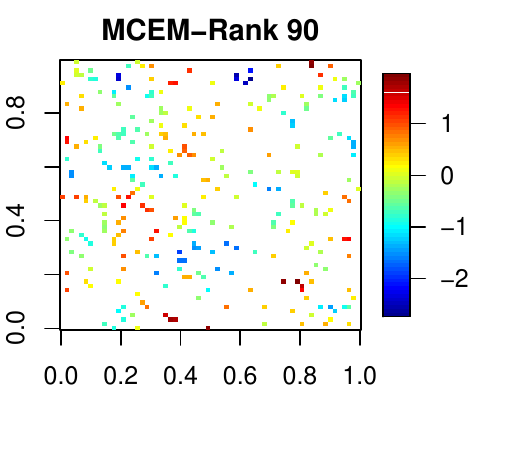}}
	\subfloat{\includegraphics[width = 0.4\linewidth,page=2]{figures/method_comp_reest_confound_quiltplot}}
	\quad
	\subfloat{\includegraphics[width = 0.4\linewidth,page=3]{figures/method_comp_reest_confound_quiltplot}}
	\subfloat{\includegraphics[width = 0.4\linewidth,page=4]{figures/method_comp_reest_confound_quiltplot}}
	\caption{Estimated random effects under the spatial-confounding case.}
	\label{fig:method_comp_reest_confound}
\end{figure}
\clearpage
}

\afterpage{%
\begin{figure}
	\centering
	\includegraphics[width = \linewidth]{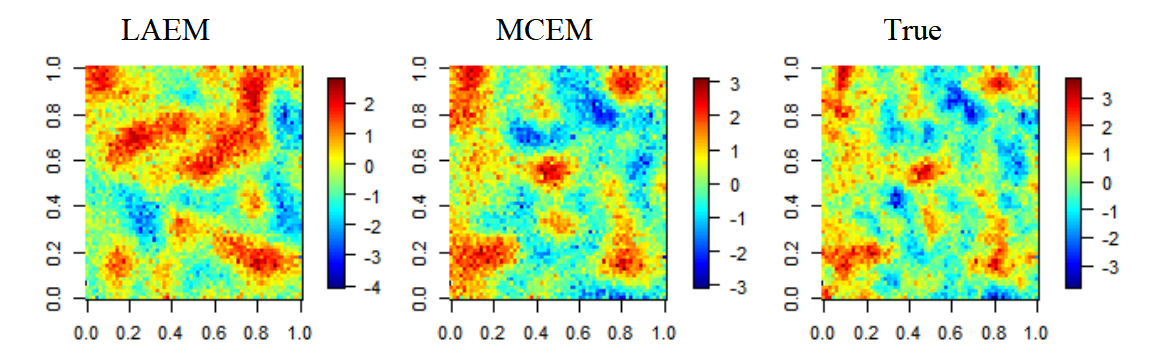}
	\caption{Estimated linear component in the conditional mean from the LAEM (left), MCEM (center) and the true (right). Data with size n=50,000 are simulated from the Mat\'{e}rn covariance model with $\nu =1.5$, $ \sigma_2=1 $ and effective range of 0.2.}
	\label{fig:massive}
\end{figure}
\clearpage
}

\end{document}